\documentstyle[aps,prl,psfig,multicol]{revtex}

\begin{document}

\def\be{\begin{eqnarray}}
\def\ee{\end{eqnarray}}
\def\nn{\nonumber}
\def\uk{u_{\bf k}}
\def\vk{v_{\bf k}}
\def\ukq{u_{{\bf k}+{\bf q}}}
\def\vkq{v_{{\bf k}+{\bf q}}}
\def\fk{f_{\bf k}}
\def\fkq{f_{{\bf k}+{\bf q}}}
\def\ek{E_{\bf k}}
\def\ekq{E_{{\bf k}+{\bf q}}}
\def\dk{\Delta_{\bf k}}
\def\dkq{\Delta_{{\bf k}+{\bf q}}}
\def\xk{\xi_{\bf k}}
\def\xkq{\xi_{{\bf k}+{\bf q}}}
\def\bfk{{\bf k}}
\def\bfq{{\bf q}}
\def\bfkq{{{\bf k} + {\bf q}}}
\def\bfkp{{{\bf k}+{\bf q}/2}}
\def\bfkm{{{\bf k}-{\bf q}/2}}
\def\xip{{\xi_+}}
\def\xim{{\xi_-}}
\def\ep{{E_+}}
\def\em{{E_-}}
\def\fp{{f_+}}
\def\fm{{f_-}}
\def\thetaq{\theta_\bfq}
\def\ident{{\bf 1}\!{\rm \bf I}}
\def\proxless{\vcenter{\hbox{$<$}\offinterlineskip\hbox{$\sim$}}}
\def\gsim{\mathrel{\rlap{\lower4pt\hbox{\hskip1pt$\sim$}}
    \raise1pt\hbox{$>$}}}

\title{Conductivity of Paired Composite Fermions}

\author{Kerwin C. Foster} 
\address
{Department of Physics, Dillard University, New Orleans, LA 70122}
\author{N.E. Bonesteel}
\address
{Department of Physics and National High Magnetic Field
Laboratory, Florida State University, Tallahassee, FL 32310}
\author{Steven H. Simon}
\address
{Bell Laboratories, Lucent Technologies, Murray Hill, NJ 07974}

\maketitle

\begin{abstract}
We develop a phenomenological description of the $\nu=5/2$ quantum
Hall state in which the Halperin-Lee-Read theory of the
half-filled Landau level is combined with a $p$-wave pairing
interaction between composite fermions (CFs).  The electromagnetic
response functions for the resulting mean-field superconducting
state of the CFs are calculated and used in an RPA calculation of
the $q$ and $\omega$ dependent longitudinal conductivity of the
physical electrons, a quantity which can be measured
experimentally.
\end{abstract}

\begin{multicols}{2}

The $\nu=5/2$ fractional quantum Hall state remains one of the
most interesting phenomena in two dimensional electron
physics\cite{ReadReview}. Since its experimental discovery over a
decade ago\cite{willett}, the nature of this state has been a
topic of debate.   Evidence from exact diagonalizations of small
systems\cite{morfandrezayi} now seems to point towards the 5/2
state being properly described as a spin-polarized Moore-Read
Pfaffian state\cite{mooreread}, a state which can be viewed as a
chiral $p$-wave superconductor\cite{mooreread,greiter} of
composite fermions (CFs) \cite{Olle}.  Among other interesting
ramifications, the Moore-Read state should theoretically exhibit
excitations with exotic nonabelian statistics\cite{mooreread} ---
something never before observed in nature.

Although there is a reasonably strong theoretical case that the
$\nu=5/2$ FQHE state is, in fact, a Moore-Read state, the question
remains, how can one test this hypothesis experimentally? While
several experiments seem to be at least {\it consistent} with the 5/2
state being a Moore-Read state\cite{Experiments,morfandrezayi}, we are
still in need of a smoking gun. The analogy with superconductivity
makes one think of how the classic experimental hallmarks of
BCS-superconductivity\cite{SchriefferBook} theory might be translated
into the fractional quantum Hall regime.  For example, in traditional
superconductors, many measurable response functions display
``coherence peaks" below the critical temperature which are extremely
good evidence of BCS superconductivity.  We would like to ask whether
such a phenomena should exist for the Moore-Read state (or, for that
matter, if any other clear signature could be seen in measurable
response functions.)  To address this question, we have developed a
phenomenological description of the FQHE state in which the
Halperin-Lee-Read(HLR) \cite{halperinleeread} theory of the
half-filled Landau level is combined with a $p$-wave pairing
interaction between CFs.  Within this theory we are able to predict
various response functions of the Moore-Read state which may be
measured experimentally.  Recalling that surface acoustic waves (SAW)
experiments\cite{Olle} were particularly powerful in experimentally
demonstrating the existence of CFs, we will be particularly interested
in the SAW signatures of the Moore-Read state.

In the HLR theory\cite{halperinleeread}, each electron in modeled as a
fermion bound to two quanta of ``Chern-Simons" flux, the fermion plus
flux being called a CF. For the 5/2 state, 4/5 of the electrons are
required to fill the lowest two (essentially inert) Landau bands and
the remaining (1/5) valence electrons are transformed to CFs. At the
mean field level, the external field precisely cancels the bound flux
and we model the valence electrons as free fermions in zero effective
magnetic field. There is some indication that under certain conditions
the residual interaction between the CFs can create a pairing
instability\cite{greiter,Bonesteel}.  To represent this physics, we
add a pairing interaction between the CFs by hand. We thus use a model
Hamiltonian for the CFs of the standard BCS form ($\hbar = c = 1$
throughout), \be
\label{eq:BCSHamiltonian} H =\!\! \sum_{{\bf k}} \xi_{\bf k}
c^\dagger_{{\bf k}} c_{{\bf k }} +\! \frac{1}{2} \! \sum_{{\bf
k},{\bf k}^\prime, {\bf q}}\!\! V_{{\bf k}{\bf k}^\prime}
c^\dagger_{{\bf k}+\frac{\bf q}{2}} c^\dagger_{-{\bf
k}+\frac{\bfq}{2}} c^{\phantom{\dagger}}_{-{\bf
k}^\prime+\frac{\bf q}{2}}
c^{\phantom{\dagger}}_{{{\bfk}^\prime+\frac{\bf q}{2}}},
\label{hgaugeinv} \ee where $c^\dagger$ is the CF creation
operator, $\xi_k = k^2/(2m) - \mu$ and $m$ is CF effective mass which
may be much larger than the underlying electron mass. (Note that the
ad-hoc mass renormalization will cause problems at the cyclotron
energy scale but is expected to be reasonable at lower
energies\cite{halperinleeread,SimonHalperin}.)  In the spirit of Ref
\onlinecite{halperinleeread} we will calculate the CF response of the
Hamiltonian (\ref{eq:BCSHamiltonian}) then transform this result (See
Eq.~\ref{eq:electron} below) to determine the physical electron
response.

In Eq.~\ref{eq:BCSHamiltonian}  the pairing interaction is taken
to be of chiral $p$-wave form $V_{\bfk \bfk^\prime} = -V e^{-i
\theta_\bfk} e^{i\theta_{\bfk^\prime}}$ where $\theta_\bfk$ is the
angle of $\bfk$ on the Fermi surface. Note that this interaction
is not time-reversal symmetric --- it is only attractive in the
$l=+1$ channel, not the $l=-1$ channel. Such an asymmetry is
expected because, although the CFs see zero average magnetic field
at the mean-field level, their residual interactions are not
time-reversal symmetric.

If we define the gap function to be
\begin{equation} \Delta_\bfq
\equiv V\sum_{\bfk^\prime} \langle c_{-\bfk^\prime+\frac{\bfq}{2}}
c_{\bfk^\prime+\frac{\bfq}{2}}\rangle e^{i\theta_{\bfk^\prime}},
\label{deltadef}
\end{equation}
the BCS mean-field Hamiltonian can be written in pseudospin notation
as
\begin{equation} H_{MF} =
\frac{1}{2} \sum_\bfk \Psi_\bfk^\dagger \left( \xi_\bfk \tau_z -
\Delta(\cos\theta_\bfk \tau_x + \sin\theta_\bfk \tau_y) \right)
\Psi_\bfk,
\end{equation}
where $\tau_x, \tau_y, \tau_z$ are the
usual Pauli spin matrices, $\Psi^\dagger_\bfk = (
c^\dagger_\bfk,~~c_{-\bfk} )$ and $\Delta = |\Delta_{\bfq = 0}|$
is the temperature dependent energy gap found by solving the usual
BCS gap equation.  It is important to note that the restriction of
$\Delta_{\bfq}$ to its zero wavevector component explicitly breaks
gauge invariance.   We will fix this problem below.

We now add a perturbation Hamiltonian to the above $H_{MF}$ given by
\begin{equation} H^\prime = \sum_{\bfq} \left[
{a_0}_\bfq  {j_0}_{-\bfq} + {a_1}_\bfq {j_1}_{-\bfq} + {a_b}_\bfq
{{j_b}_{-\bfq}} \right],
\end{equation}
where
\begin{eqnarray}& & \!\!\! ({j_0}_\bfq,{j_1}_\bfq) = \frac{1}{2} e \sum_{\bfk}
\Psi^\dagger_{\bfkp} (\tau_z,\frac{k_\perp}{m}\tau_0) \Psi_\bfkm,
\\  {j_b}_\bfq &=& \frac{1}{2}\sum_k \Psi_\bfkp^\dagger
(-\cos\theta_\bfk\tau_y + \sin\theta_\bfk \tau_x) \Psi_\bfkm.
\end{eqnarray}
The first two terms in $H'$ are the coupling of the scalar potential
${a_0}_\bfq$ to the density ${j_0}_\bfq$ and the transverse vector
potential ${a_1}_\bfq$ to the transverse paramagnetic current
${j_1}_\bfq$(for simplicity we work in Coulomb gauge here so the
longitudinal vector potential is zero).  The third term in $H'$ is the
coupling of CFs to the phase fluctuations of the order parameter,
described by ${a_b}_\bfq = (\Delta_\bfq-\Delta_{-\bfq}^*)/(2i)$ which
will be self-consistently calculated.  Such a self-consistent
treatment of phase fluctuations is a standard
method\cite{andersonrickayzen} that enables one to calculate gauge
invariant responses to external perturbations despite the fact that
$H_{MF}$ is not gauge invariant by itself.  Magnitude fluctuations are
neglected since they can be shown to decouple due to approximate
particle-hole symmetry at the Fermi surface \cite{kulik}.

We define the response functions $Q_{ij}$ for the mean-field
Hamiltonian $H_{MF}$ by 
\be j_i(\bfq,\omega) &=& Q_{ij}(\bfq,\omega)
a_j({\bf q},\omega), \ee 
where the indices $i$ and $j$ can be $0$, $1$
or $b$.  Here $j_i(\bfq,\omega)$ is the fourier transform of the
time-dependent expectation value of ${j_i}_\bfq + \delta_{i1} (n/m)
{a_1}_\bfq$ and thus includes the diamagnetic contribution to the
transverse current.

When the constraint, following from (\ref{deltadef}), that ${a_b}_\bfq
= V \langle {j_b}_\bfq\rangle$ is included, the standard RPA analysis
\cite{andersonrickayzen} can be used to obtain the gauge invariant CF
electromagnetic response functions $K_{\mu\nu}$ defined by
\begin{equation} j_\mu(\bfq,\omega) = K_{\mu\nu}(\bfq,\omega)a_\nu({\bf
q},\omega),
\end{equation}
where the indices $\mu$ and $\nu$ can now be $0$ or $1$. We obtain
\begin{equation} K_{\mu\nu} = Q_{\mu\nu}
- Q_{\mu b} Q_{b \nu}/(Q_{bb} - 1/V),
\end{equation}
where the second term on the right hand side corresponds to the
usual vertex corrections required for a conserving approximation.

Following Mattis and Bardeen \cite{mattisbardeen} (see also
\cite{ubbenslee}), in the extreme anomalous limit $v_F q \gg {\rm
Max}[\omega,\Delta]$ the expressions for these response functions
can be simplified substantially.  We obtain
\begin{eqnarray}
\label{eq:firstQ} Q_{00}(\bfq,\omega) &=& -\frac{m^2}{4\pi^2}
F_0(q) \Omega(1,-\cos\theta_q,\omega) -
\frac{m}{2\pi},\\
Q_{10}(\bfq,\omega) &=& i \sin\thetaq \frac{m}{4\pi^2} F_1(q)
\Omega(0,1,\omega), \\
Q_{11}(\bfq,\omega) &=& \frac{1}{4\pi^2} F_2(q)
\Omega(1, \cos\theta_q,\omega) - \frac{q^2}{24\pi m},\\
 Q_{b0}(\bfq,\omega) &=& i \frac{m^2}{4\pi^2} F_0(q)
\frac{\omega}{\Delta} \cos\frac{\theta_\bfq}{2}\Omega(0,1,\omega),
\\
Q_{b1}(\bfq,\omega) &=& \frac{m}{4\pi^2} F_1(q)
\frac{\omega}{\Delta} \sin\frac{\theta_\bfq}{2}\Omega(0,1,\omega),
\\ Q_{bb}(\bfq,\omega)  &=&
\frac{m^2}{4\pi^2}F_0(q)\Omega(1,-1,\omega) + \Lambda(\bfq) +
\frac{1}{V}, \label{eq:lastQ}
\end{eqnarray}
where $\theta_q$ is the angle between the vectors $\bfkp$ and
$\bfkm$ when constrained to the Fermi surface, and
$Q_{0b}(\bfq,\omega) = -Q_{b0}(\bfq,\omega)$,
$Q_{b1}(\bfq,\omega)= Q_{1b}(\bfq,\omega)$,  $Q_{10}(\bfq,\omega)=
Q_{01}(\bfq,\omega)$.  In these equations, \be F_\alpha(q) = (2
k_f^{\alpha-1}/q) \left[ 1 - q^2/(2 k_f)^2\right]^{(\alpha-1)/2},
\ee and  $\Omega(\omega) = \Omega_1(\omega) + i \Omega_2(\omega)$
with
\end{multicols}
\be \Omega_1(r,s,\omega) &=& \pi
\int_{{\rm max}[\Delta-\omega,-\Delta]}^\Delta \left(1-2f(E+\omega)\right)
\frac{sE(E+\omega)+r\Delta^2} {[\Delta^2-E^2]^{1/2}
[(E+\omega)^2-\Delta^2]^{1/2}} dE, \\ \nn \\ \Omega_2(r,s,\omega)
&=& -2 \pi \int_\Delta^\infty \left(f(E)-f(E+\omega)\right)
\frac{sE(E+\omega)+r\Delta^2}{[E^2-\Delta^2]^{1/2}[(E+\omega)^2-\Delta^2]^{1/2}}dE
\nn\\ &&~~~~~~~
-\pi \int_{\Delta-\omega}^{-\Delta}
\left(1-2f(E+\omega)\right)
\frac{sE(E+\omega)+r\Delta^2}{[E^2-\Delta^2]^{1/2}[(E+\omega)^2-\Delta^2]^{1/2}}dE,
\\ \Lambda(\bfq) &=& \int\frac{d^2 k}{(2\pi)^2} \left(
\frac{1-f(\xi_{\bfkm}) - f(\xi_{\bfkp})}{\xi_{\bfkm} +
\xi_{\bfkp}} - \frac{1-2f(E_\bfk)}{2E_\bfk} \right),
\ee
\begin{multicols}{2}
\noindent where $E_\bfk = \sqrt{\xi_\bfk^2 + |\Delta|^2}$ and $f$
is the Fermi function. The one dimensional integrals for
$\Omega(\omega)$ are easily evaluated numerically. In the extreme
anomalous limit $\Lambda(\bfq) \simeq - (m/(2 \pi)) \log(v_F q/(2
\Delta(0))$ is large and the vertex corrections to the Coulomb
gauge are small.

Note that this mean-field treatment gives a finite temperature
phase transition.  It should be emphasized that this is an
artifact of our calculation.  Vortices in a Chern-Simons
``superfluid'' cost a finite amount of energy to create and
interact only via short-range interactions.  As a result there is
no finite temperature Kosterlitz-Thouless transition and
fluctuations will push $T_c$ to zero.  We assume here that
including these fluctuations will primarily have the effect of
smoothing the finite temperature transition into a crossover, but
the qualitative features of our results will remain.

\begin{figure}[t]
\centerline{\psfig{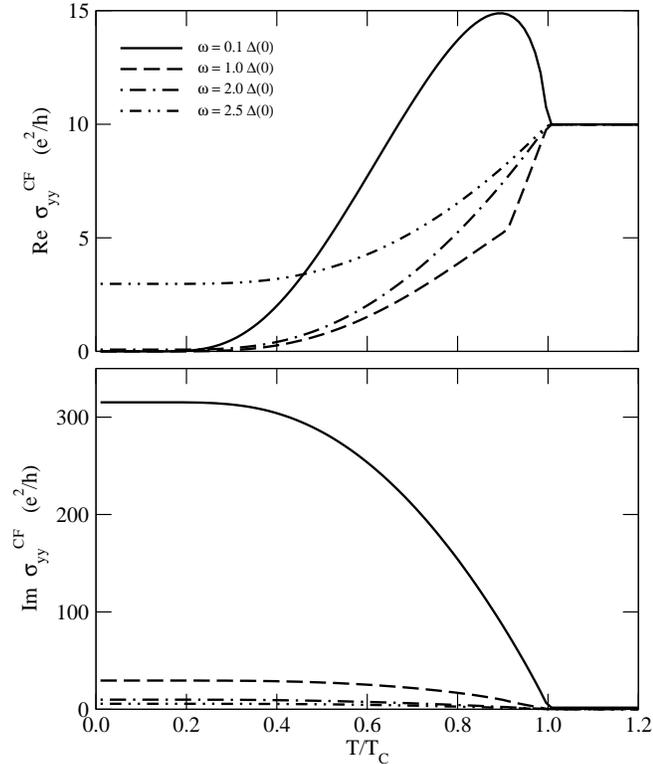}}
\caption{Real and Imaginary parts of the transverse conductivity of
composite fermions, $\sigma_{yy}^{CF}$, in a $p$-wave
``superconducting'' state as a function of temperature for
$\omega/\Delta(0) =$ 0.1, 1.0, 2.0 and 2.5.  For low frequencies,
$\omega\ \proxless\ 0.2 \Delta$, ${\rm Re}\ \sigma_{yy}^{CF}$ shows a
Hebel-Slichter coherence peak and ${\rm Im}\ \sigma_{yy}^{CF}$ shows a
strongly enhanced diamagnetic response.  Results are for $\Delta(0) =
0.01 E_F$ and $q = 0.1 k_F$.}
\label{sigmayy_cf}
\end{figure}

To fix the parameters of our model, in all of what follows we take
$\Delta(0)/E_F = 0.01$.  This is consistent with $k_f = (4\pi
n/5)^{1/2} \sim 10^8$ m$^{-1}$, $\Delta(0) \sim 0.1$ K, and a CF
effective mass $m \sim 10 m_b$ where $m_b$ is the electronic band
mass.

\vskip .1in

\begin{figure}[t]
\centerline{\psfig{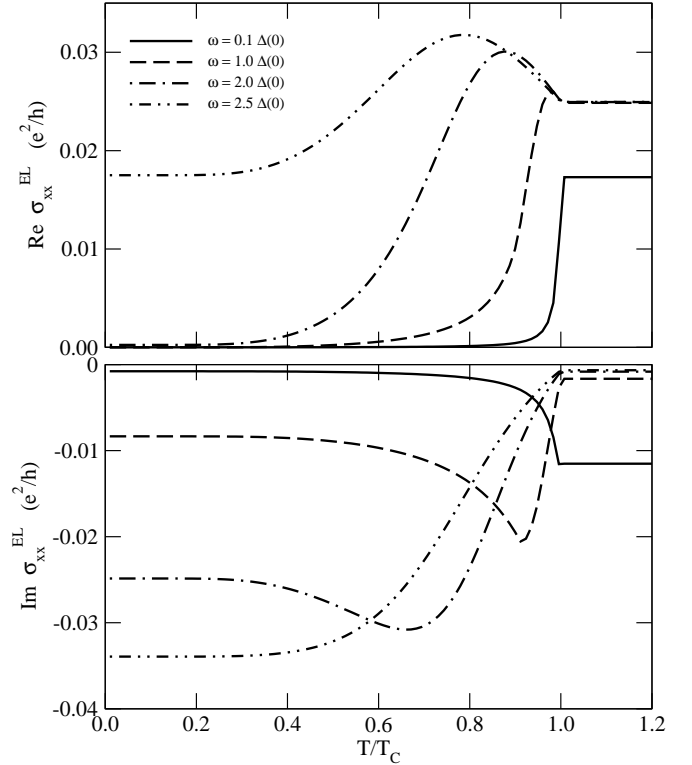}}
\caption{Real and imaginary parts of the longitudinal conductivity of
electrons, $\sigma_{xx}^{EL}$, for the same parameters as Fig.~1.  In
this regime, $\sigma_{xx}^{EL} \propto 1/\sigma_{yy}^{CF}$, and
because of the strongly enhanced ${\rm Im}\ \sigma_{yy}^{CF}$ for
$\omega\ \proxless\ 0.2 \Delta$ (see Fig.~1) there is no sign of the
Hebel-Slichter peak in $\sigma_{xx}^{EL}$.} \label{sigmaxx_el_1}
\end{figure}

Coherence effects are most clearly seen in the $q$ and $\omega$
dependent conductivity.  Figure \ref{sigmayy_cf} shows the transverse
conductivity $\sigma^{CF}_{yy} = e^2 K_{11}/i\omega$ for CFs as a
function of temperature for $q = 0.1 k_F$.  For low frequencies,
$\omega\ \proxless\ 0.2 \Delta(0)$, ${\rm Re}\ \sigma^{CF}_{yy}$ shows
a Hebel-Slichter coherence peak just below $T_c$.  This peak appears
because for small $q$ the $p$-wave nature of the pairing is irrelevant
and the coherence factors which determine electromagnetic absorption
are Type II, the same coherence factors which govern the temperature
dependence of the NMR relaxation time $1/T_1$ in conventional
superconductors.  For the same low frequencies ${\rm Im}\
\sigma^{CF}_{yy}$ increases dramatically below $T_c$, reflecting the
large increase in ${\rm Re}\ K_{11}(q,\omega)$ due to the enhanced CF
diamagnetic response in the paired state.  Note that the kink clearly
visible in ${\rm Re}\ \sigma^{CF}_{yy}$ for $\omega = \Delta(0)$
occurs when the threshold condition $\omega = 2 \Delta(T)$ is
satisfied.

It is natural to ask if a similar coherence peak is observable in the
5/2 state.  To address this we calculate the experimentally measurable
{\it electronic} longitudinal conductivity, $\sigma_{xx}^{EL}$,
following HLR using the Chern-Simons RPA.  The only modification to
the HLR result is due to the off-diagonal part of the mean-field CF
response function -- a consequence of the chiral nature of the
$p$-wave state.  The resulting expression for the conductivity is
\begin{equation} \sigma^{EL}_{xx} =
 \frac{i e^2 \omega K_{00}/q^2}{1+\frac{4\pi i \tilde \phi}{q} K_{10}
+ \frac{(2\pi\tilde\phi)^2}{q^2} (K_{00} K_{11} - K_{10}^2)},
\label{eq:electron} \end{equation} where $\tilde \phi = 2$ is the
number of flux quanta attached to each CF. Just as in the HLR case, in
the limit of small $q$ this expression is dominated by $K_{11}$ and to
a good approximation $\sigma^{EL}_{xx} \simeq (e^2/(2\pi\tilde\phi)^2)
i\omega K_{11}^{-1} = (e^2/(2\pi \tilde\phi))^2/\sigma_{yy}^{CF}$.

Figure \ref{sigmaxx_el_1} shows the electronic longitudinal
conductivity for the same parameters as Fig.~\ref{sigmayy_cf}. The
main observation is that there is no sign of the Hebel-Slichter peak
at low frequencies.  This is because of the rapid increase in ${\rm
Im}\ \sigma_{yy}^{CF}$ below $T_c$ discussed above.  This rapid
increase suppresses $\sigma_{xx}^{EL}$ below $T_c$, masking the
relatively small Hebel-Slichter peak.  We note that if one could
measure the real and imaginary parts of $\sigma_{xx}^{EL}$ to
sufficient accuracy to carry out the inversion to obtain
$\sigma_{yy}^{CF}$ one could in principle observe the Hebel-Slichter
peak, although in practice such accuracy would be very difficult to
achieve.

Note that for $\omega \gsim \Delta $ a peak in $\mbox{Re}\
\sigma_{xx}^{EL}$ does appear below $T_C$.  We emphasize that this is
not a coherence peak but rather a consequence of the fact that the
absolute magnitude of $\sigma_{yy}^{CF}$ decreases below $T_c$ for
these frequencies.

\vskip .1in

\begin{figure}[t]
\centerline{\psfig{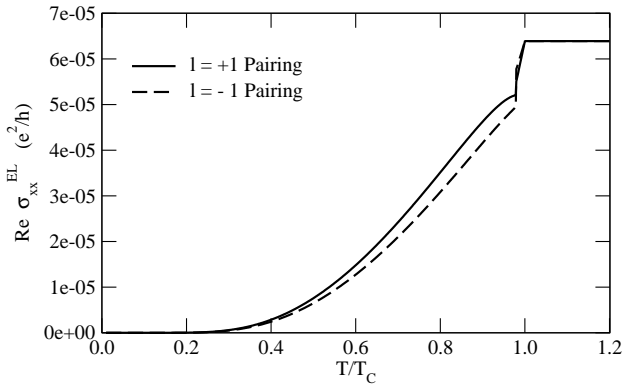}}
\vskip .15in
\caption{Longitudinal conductivity of electrons for $\omega = 0.5 \Delta(0)$
and $q = 0.75 k_F$.  Results are shown for pairing angular momentum parallel
and antiparallel to the applied field.  The small difference indicates the
$p$-wave nature of the pairing is difficult to observe in $\sigma^{EL}_{xx}$, even
at large wavevectors.  Results are for $\Delta(0) = 0.01 E_F$.}
\label{pwaveness}
\end{figure}

All the results shown to this point are for $q \ll k_F$.  This is
the regime for which the HLR theory is expected to be
qualitatively correct.  It must be emphasized that in this limit
the $p$-wave nature of the pairing is irrelevant and the results
would be the same for $s$-wave (up to factors of 2 from the fact
that we need two spin states), or any $l$-wave, CF
superconductors. The $p$-wave nature of the pairing only becomes
relevant when $q$ is large enough to span parts of the Fermi
surface where the phase of the order parameter is significantly
different.

A measure of the relevance of the $p$-wave pairing can be seen by
comparing results for which the applied magnetic field is parallel and
antiparallel to the pair angular momentum.  This corresponds to
changing the sign of $\tilde\phi$ in (\ref{eq:electron}).  For $q \ll
k_F$, including all results presented above, there is no measurable
difference for these two cases.  For $q \sim k_F$, a difference in
$\sigma^{EL}_{xx}$ appears, but it is small.  A typical result is
shown in Fig.~\ref{pwaveness}.

To summarize, we have developed a phenomenological model of the 5/2
state by adding a chiral $p$-wave pairing interaction between CFs by
hand.  The electromagnetic CF response functions for this model were
then calculated, including self-consistent fluctuations of the order
parameter to ensure gauge invariance.  For small $q$ the CF transverse
conductivity exhibits a Hebel-Slichter peak, but this peak is not
easily observable in measurements of the electronic longitudinal
conductivity.  Although we have focused on the question of whether
clear signatures of superconductivity can be seen in SAW measurements,
similar calculations can give predicitions for other electromagnetic
response experiments, such as microwave conductivity and resonant
Raman scattering\cite{dassarma}. Furthermore, the methods described here can
be more generally applied to analyze a variety of other paired CF
states --- including the Haldane-Rezayi\cite{haldanerezayi} state and several
proposed paired bilayer states\cite{paired}.

The authors thank Kun Yang for useful discussions. KCF and NEB
acknowledge support from US DOE Grant No.\ DE-FG02-97ER45639.

\end{multicols}

\end{document}